\documentclass[prl,aps,twocolumn]{revtex4}

\usepackage{epsfig} 
\usepackage{graphics}

\begin{document}

\title{Volume-energy correlations in the slow degrees of freedom of computer-simulated phospholipid membranes}

\author{Ulf R. Pedersen,$^*$ G\"{u}nther H. Peters,$^\dagger$ Thomas B. Schr\o{}der$,^*$ and Jeppe C. Dyre$^*$}

\affiliation{$^*$DNRF Centre ``Glass and Time,'' IMFUFA, Department of Sciences,
Roskilde University, Postbox 260, DK-4000 Roskilde, Denmark\\$^\dagger$Center for Membrane Biophysics (MEMPHYS), Department of Chemistry, Technical University of Denmark, DK-2800 Lyngby, Denmark}


\begin{abstract}
Constant-pressure molecular-dynamics simulations of phospholipid membranes in the fluid $L_\alpha$ phase reveal strong correlations between equilibrium fluctuations of volume and energy on the nanosecond time-scale. The existence of strong volume-energy correlations was previously deduced indirectly by Heimburg from experiments focusing on the phase transition between the $L_\alpha$ and the $L_\beta$ phases. The correlations, which are reported here for three different membranes (DMPC, DMPS-Na, and DMPSH), have volume-energy correlation coefficients ranging from 0.81 to 0.89. The DMPC membrane was studied at two temperatures showing that the correlation coefficient increases as the phase transition is approached.
\end{abstract}

\maketitle

Biological membranes are essential parts of living cells. They not only act as passive barriers between outside and inside, but also play an active role in various biological mechanisms. The major constituent of biological membranes are phospholipids. Pure phospholipid membranes often serve as a models for the more complex biological membranes. Close to physiological temperatures membranes undergo a transition from the high-temperature fluid $L_\alpha$ phase (often referred to as the ``biologically relevant phase'') to a low-temperature ordered gel phase $L_\beta$. In the melting regime response functions such as heat capacity, volume-expansion coefficient, and area-expansion coefficient increase dramatically. Also, the characteristic time for the collective degrees of freedom increases and becomes longer than milliseconds. Some time ago Heimburg found that the slow, dominating parts of heat capacity and volume-expansion coefficient of DMPC as a function of temperature can be superimposed close to the melting temperature $T_m$ \cite{Heimburg1998} (see also Refs. 2 and 3). Thus the response functions are related in such a way that a single function describes the temperature dependence of both.

The fluctuation-dissipation (FD) theorem connects (linear) response functions to equilibrium fluctuations. The isobaric heat capacity $c_p$ can be calculated from enthalpy fluctuations as follows: $c_p=\langle (\Delta H)^2\rangle/(Vk_BT^2)$, where $\langle \ldots \rangle$ is an average in the constant temperature and pressure ensemble and $\Delta$ is deviation from the average value. Similarly, volume fluctuations are connected to the isothermal volume compressibility by the expression $\kappa_T=\langle (\Delta V)^2\rangle/(Vk_BT)$. If the response functions were described by a single parameter, fluctuations are also described by a single parameter \cite{Heimburg1998,Ellegaard2007} and the microstates were connected via the relation $\Delta H_i=\gamma^{vol}\Delta V_i$. At constant pressure this relation applies if and only if $\Delta E_i=\gamma^{vol}\Delta V_i$ (where $E$ is energy), which is the relation investigated below. This situation is referred to as a that of a single-parameter description \cite{Ellegaard2007}. A single-parameter description applies to a good approximation for several models of van der Waals bonded liquids as well as for experimental super-critical argon \cite{Pedersen2006b}.

Unfortunately, molecular-dynamics simulations are not possible for investigating ``single parameter''-ness of membranes close to $T_m$, because the relaxation time for the collective modes by far exceeds possible simulation times. We show below, however, that a single-parameter description applies to a good approximation for the slow degrees of freedom of the fluid L$_\alpha$ phase, a description that applies better upon approaching $T_m$. At the end of this note we briefly discuss how this property may be tested in experiments monitoring frequency-dependent thermoviscoelastic response functions.

It is not {\it a priori} obvious that a single parameter may be sufficient for describing slow thermodynamic fluctuations of a membrane. For instance, simulations of water and methanol showed no ``single parameter''-ness. Apparently, what happens here is that contributions to volume and energy fluctuations from hard-core repulsion compete with those from directional hydrogen bonds to destroy any significant correlation \cite{Pedersen2006b}. Membranes are complex anisotropic systems, and we cannot give any obvious reason that volume and energy should correlate strongly in their fluctuations.

A membrane may be pertubated via three thermodynamic energy bonds. The change of enthalpy $dH$ can be written as a sum of contributions from a thermal energy bond, a mechanical volume energy bond, and a mechanical area energy bond, $dH=dE+pdV+\Pi dA$, where $p$ is pressure, $V$ volume, $\Pi$ membrane surface tension, and $A$ membrane area. The natural ensemble to consider is the constant $T$, $p$, and $\Pi$ ensemble, since the surrounding water acts as a reservoir. If a single parameter controls the microstates, for all states $i$ one would have $\Delta E_i=\gamma^{vol}\Delta V_i=\gamma^{area}\Delta A_i$ where the $\gamma$'s are constants. In general, the microstates may of course be controlled by several parameters. An interesting question is how many parameters are sufficient to describe the membrane thermodynamics to a good approximation. This question is addressed below by investigating molecular-dynamics simulations of different phospholipid membranes.


An overview of the simulated systems is found in table \ref{table1}. The simulated systems include different head groups (both charged and zwitterions) and temperatures. All simulations was carried out in the constant pressure, temperature ensemble. The membranes are fully hydrated and in the fluid $L_\alpha$ phase. The simulations were performed using the program NAMD \cite{Phillips2005} and a modified version of CHARMM27 all hydrogen parameter set \cite{Foloppe2000,Pedersen2006}. More simulation details are given in Ref. 6.

\begin{table}[b!]
\begin{tabular}{lccccc}
\hline
  & {$T$ [K]}
  & {$R_{EV}$}
  & {$A_{lip}$ [\AA{}$^2$]}
  & {$t$ [ns]}
  & {$t_{tot}$ [ns]}
 \\
\hline
DMPC  & 310  & 0.885  & 53.1 & 60 & 114  \\
DMPC  & 330  & 0.806  & 59.0 & 50 & 87   \\
DMPS-Na  & 340  & 0.835  & 45.0 & 22 & 80   \\
DMPSH & 340  & 0.826  & 45.0 & 40 & 77   \\
\hline
\end{tabular}
\caption{Data from simulations of fully hydrated phospholipid membranes of 1,2-Dimyristoyl-sn-Glycero-3-Phosphocholine (DMPC),  1,2-Dimyristoyl-sn-Glycero-3-Phospho-L-Serine with sodium as counter ion (DMPS-Na) and hydrated DMPS (DMPSH). The columns list temperature, volume-energy correlation coefficient, average lateral area per lipid, simulation time in equilibrium (used in the data analysis), and total simulation time.}
\label{table1}
\end{table}

\begin{figure}[!t]
\includegraphics[width=\columnwidth]{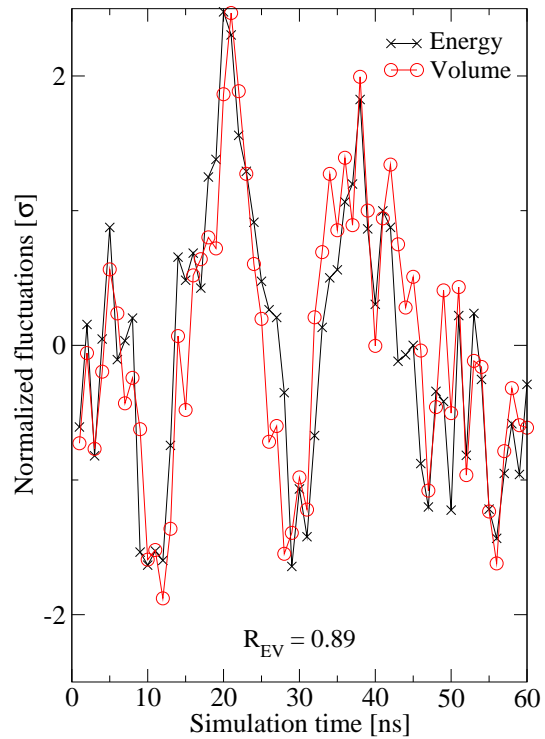}
\caption{\label{pedersen:1} Normalized fluctuations of energy ($\times$) and volume ($\circ$) for a DMPC membrane at 310 K. Each data point represents a 1 ns average. Energy and volume correlate with correlation coefficient $R_{EV}=0.89$.}
\end{figure}

\begin{figure}[!t]
\includegraphics[width=\columnwidth]{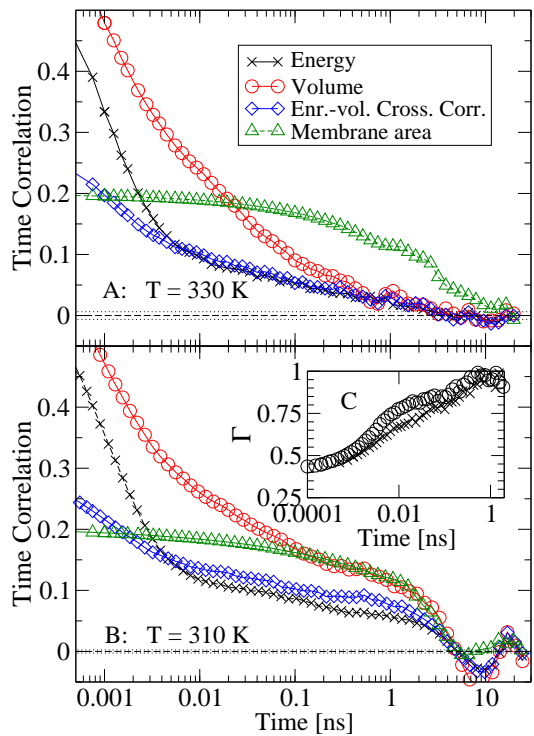}
\caption{\label{pedersen:2} Time-correlation functions for a DMPC membrane of potential energy $C_{EE}$ ($\times$), total volume $C_{VV}$ ($\circ$), membrane area $C_{AA}$ ($\bigtriangleup$), and cross correlation between energy and volume $C_{EV}$ ($\diamond$). Time correlation for membrane areas are scaled by a factor 0.2. Panel A shows data at 330 K, panel B at 310 K. The inset (C) displays $\Gamma(t)=C_{EV}(t)/\sqrt{C_{EE}(t)C_{VV}(t)}$ at 310 K ($\circ$) and 330 K ($\times$). $\Gamma$ approaches unity at $t\simeq1$ ns showing that volume and energy correlate strongly on this timescale.}
\end{figure}

The correlations between equilibrium time-averaged fluctuations of volume and energy on the nanosecond time-scale of a DMPC membrane at 310 K are shown on Fig. \ref{pedersen:1}. If $\overline{E(t)}$ and $\overline{V(t)}$ is the energy and volume averaged over 1 nanosecond, the figure shows that to a good approximation one has
\begin{equation}
\Delta\overline{E(t)}\simeq\gamma^{vol}\Delta\overline{V(t)}
\label{eq1}
\end{equation}
where $\gamma^{vol}=\sigma_E/\sigma_V$ is a constant (standard deviation $\sigma$) and $\Delta$ is difference from the thermodynamical average value. Similar results were found for the other membranes studied. Table \ref{table1} shows that volume-energy correlation coefficients ($R_{EV}={\langle\Delta\overline{E}\Delta\overline{V}\rangle}/\sqrt{\langle(\Delta\overline{E})^2\rangle\langle(\Delta\overline{V})^2\rangle}$) range between 0.81 and 0.89.

The correlation depends on the time scales considered. This can be investigated by evaluating $\Gamma(t)=C_{EV}(t)/\sqrt{C_{EE}(t)C_{VV}(t)}$ where $C_{AB}(t)=\langle \Delta A(\tau)\Delta B(\tau+t)\rangle/\sqrt{\langle (\Delta A)^2\rangle\langle (\Delta B)^2\rangle}$ is a time correlation function. $\Gamma(t)=0$ implies that energy at time $\tau$ is uncorrelated with volume at time $t+\tau$, whereas $\Gamma(t)$ close to unity implies strong correlation. $\Gamma(t)$ is plotted in the inset of Fig. \ref{pedersen:2}. At short time (picoseconds) $\Gamma$ is around 0.5 but approaches unity at $t\simeq1$ ns.

The ``single-parameter''-ness between volume and energy is closely connected to the experimental findings of Heimburg \cite{Heimburg1998}, since the slow (collective) degrees of freedom fluctuating on time scales larger than $1$ ns are those giving rise to the dramatic changes of the response functions approaching $T_m$. Figure \ref{pedersen:2} shows the time-correlation functions of energy, volume, and area of a DMPC membrane at 330 K and 310 K. Time constant as well as magnitude of the slow fluctuations increase when temperature decreases and the phase transition is approached. $\gamma^{vol}$ in Eq. (\ref{eq1}) is $9.3\times10^{-4}$ cm$^3$/J. This is of the same order of magnitude as $\gamma^{vol}=7.7\times10^{-4}$ cm$^3$/J calculated from the experimental data of $C_p(T)$ and $\kappa_T^{vol}(T)$ at $T_m$ \cite{Heimburg1998}.

Both volume and energy time-correlation functions show a two-step relaxation at 310 K for DMPC (Fig. \ref{pedersen:2}B). As temperature is lowered towards $T_m$, the separation is expected to become more significant. It makes sense to divide the dynamics into two separated processes, a fast and a slow collective process. Our simulations suggest that the slow degrees of freedom can be described by a single parameter, but not the fast degrees of freedom.

To see the ``single-parameter''-ness of the $L_\alpha$ phase, the fast degrees of freedom must be filtered out. Experiments deal with macroscopic samples where fluctuations are small and difficult to measure (the relative magnitude of fluctuations goes as $1/\sqrt{N}$ where $N$ is the number of molecules). It is therefore difficult in experiment to perform the same analysis as we have done here; it is easier to measure response functions. Fast fluctuations can be filtered out by measuring frequency-dependent response functions. The slow collective degrees of freedom give rise to a separate ``loss'' peaks in the imaginary parts. A frequency-dependent Prigogine-Defay ratio $\Lambda_{Tp}(\omega)$ was recently suggested as a test quantity for single-parameter-ness in a paper focusing on the properties of glass-forming liquids \cite{Ellegaard2007}. If $c_p''(\omega)$, $\kappa_T''(\omega)$, and $\alpha_p''(\omega)$ are the imaginary parts of the frequency-dependent isobaric specific heat (per volume), isothermal compressibility, and isobaric expansion coefficient, respectively, by definition

\begin{equation}
 \Lambda_{Tp}(\omega)=\frac{c_p''(\omega)\kappa_T''(\omega)}{T[\alpha_p''(\omega)]^2}.
\end{equation}
In general $\Lambda_{Tp}(\omega)\geq1$, and $\Lambda_{Tp}(\omega)=1$ if and only if a single parameter describes the fluctuations \cite{Ellegaard2007}. The quantity $1/\sqrt{\Lambda_{Tp}}$ may be interpreted as a correlations coefficient.

In summary, we found strong volume-energy correlations of the slow degrees of freedom in molecular-dynamics simulations of different phospholipid membranes in the $L_\alpha$ phase. An experimental test was suggested.

\section{Acknowledgments}
 The authors would like to thank Nick P. Bailey, Thomas Heimburg, and S\o{}ren Toxv\ae{}rd for fruitful discussions and useful comments.
This work was supported by the Danish National Research Foundation Centre for Viscous Liquid Dynamics ``Glass and Time.``
GHP acknowledges financial support from the Danish National Research Foundation via a grant to the MEMPHYS-Center for Biomembrane Physics. 
Simulations were performed at the Danish Center for Scientific Computing at the University of Southern Denmark.

\end{document}